\documentclass{article}

\usepackage{microtype}
\usepackage{graphicx}
\usepackage{subfigure}
\usepackage{multirow}
\usepackage{booktabs} 

\usepackage{hyperref}

 \usepackage[accepted]{icml2025}

\usepackage{amsmath}
\usepackage{amssymb}
\usepackage{mathtools}
\usepackage{amsthm}

\usepackage[capitalize,noabbrev]{cleveref}

\theoremstyle{plain}

\theoremstyle{definition}

\theoremstyle{remark}

\usepackage[textsize=tiny]{todonotes}

\icmltitlerunning{}

\begin{document}

\twocolumn[
\icmltitle{Transformer-Enhanced Variational Autoencoder for \\Crystal Structure Prediction}



\icmlsetsymbol{equal}{*}

\begin{icmlauthorlist}
\icmlauthor{Ziyi Chen}{yyy,sch}
\icmlauthor{Yang Yuan}{yyy,sch}
\icmlauthor{Siming Zheng}{comp1}
\icmlauthor{Jialong Guo}{yyy}
\icmlauthor{Sihan Liang}{yyy,sch}
\icmlauthor{Yangang Wang}{yyy,sch}
\icmlauthor{Zongguo Wang}{yyy,sch}
\end{icmlauthorlist}

\icmlaffiliation{yyy}{Computer Network Information Center, Chinese Academy of Sciences, Beijing, China}
\icmlaffiliation{comp1}{Siming Zheng is with vivo Mobile Communication Co., Ltd, Hangzhou, China}

\icmlaffiliation{sch}{University of Chinese Academy of Sciences, Beijing, China}

\icmlcorrespondingauthor{Zongguo Wang}{wangzg@cnic.cn}

\icmlkeywords{Machine Learning, ICML}

\vskip 0.3in
]



\printAffiliationsAndNotice{\icmlEqualContribution} 

\begin{abstract}
Crystal structure forms the foundation for understanding the physical and chemical properties of materials. Generative models have emerged as a new paradigm in crystal structure prediction(CSP), however, accurately capturing key characteristics of crystal structures, such as periodicity and symmetry, remains a significant challenge. In this paper, we propose a Transformer-Enhanced Variational Autoencoder for Crystal Structure Prediction (TransVAE-CSP), who learns the characteristic distribution space of stable materials, enabling both the reconstruction and generation of crystal structures. TransVAE-CSP integrates adaptive distance expansion with irreducible representation to effectively capture the periodicity and symmetry of crystal structures, and the encoder is a transformer network based on an equivariant dot product attention mechanism. Experimental results on the carbon\_24, perov\_5, and mp\_20 datasets demonstrate that TransVAE-CSP outperforms existing methods in structure reconstruction and generation tasks under various modeling metrics, offering a powerful tool for crystal structure design and optimization.

\end{abstract}

\section{Introduction}
\label{introduction}

Crystal structure prediction (CSP) methods play a crucial role in the discovery of novel materials and remain a persistent challenge in condensed matter physics and materials chemistry. With the advancement of first-principles calculations, such as Density Functional Theory (DFT)\cite{hohenberg1964inhomogeneous} and structure search algorithms, including simulated annealing \cite{kirkpatrick1983optimization} and genetic algorithms\cite{wu2013adaptive}, the precision and efficiency of CSP have improved significantly. These methods typically involve generating candidate crystal structures and performing DFT calculations to screen them. Notable examples include USPEX\cite{lyakhov2013new} and AIRSS\cite{pickard2011ab}. However, these approaches are computationally intensive and time-consuming. Traditional computationally driven CSP methods often lack effective learning priors and are generally limited to structural searches within a single system. Furthermore, due to the inherent randomness of the search space, they require substantial computational resources and are susceptible to getting trapped in local optima.

With the rapid advancement of deep learning technologies, CSP has progressed into a new phase of data-driven. Neural network models now substitute DFT calculations to predict material properties\cite{merchant2023scaling}, significantly enhancing the efficiency of structure searches. Generative models, including variational autoencoder\cite{kingma2013auto}, adversarial generative networks\cite{goodfellow2014generative}, and diffusion models\cite{ho2020denoising} with prior knowledge, are increasingly used in crystal structure generation. These models are capable of learning the feature space distribution of sample crystal structures and directly sampling from this space to generate crystal structures that adhere to the learned distribution. This approach significantly improves search efficiency compared to traditional CSP methods.

\begin{figure*}[ht]
\vskip 0.2in
\begin{center}
\centerline{\includegraphics[width=(\columnwidth*2)]{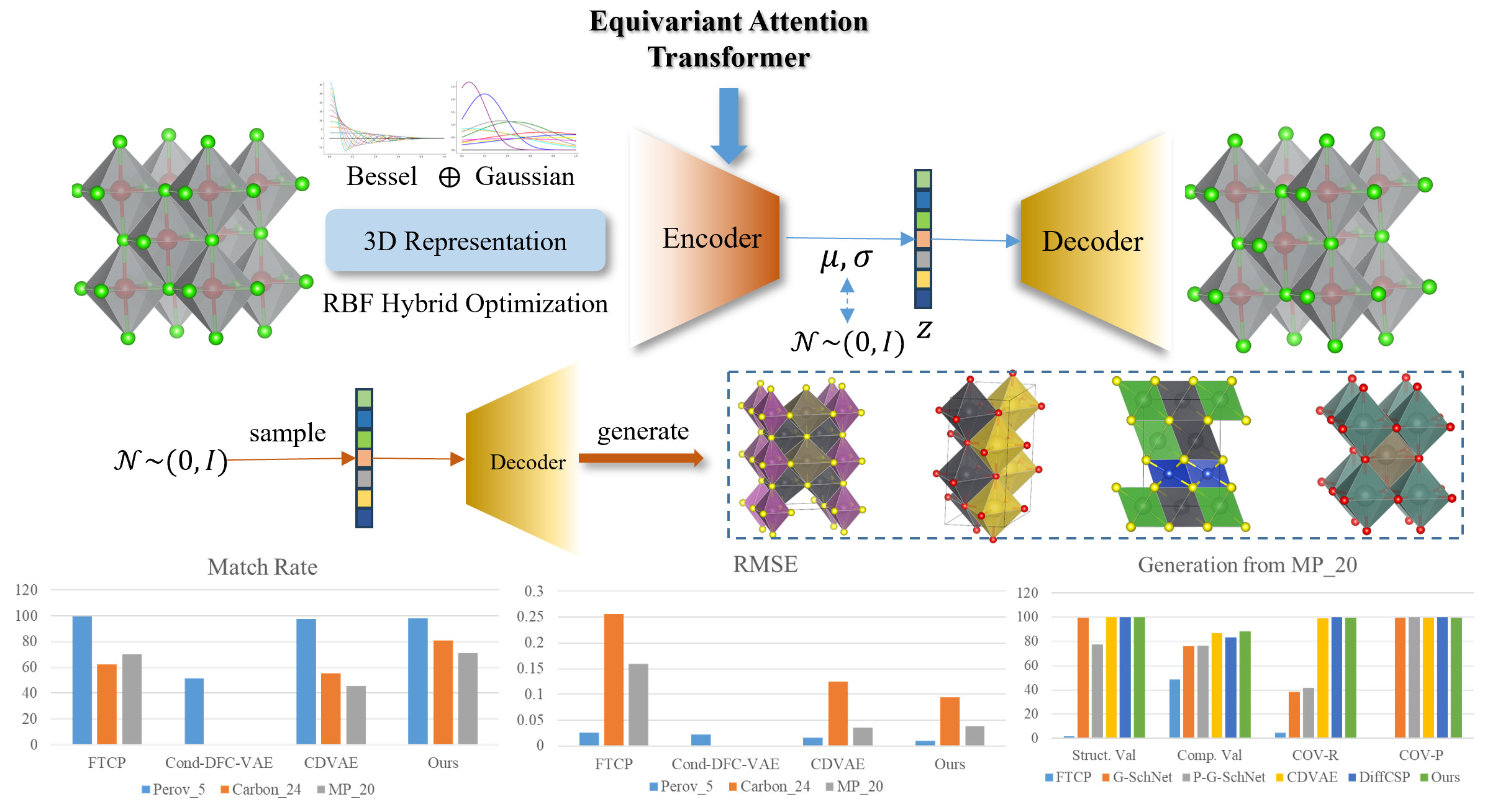}}
\caption{TransVAE-CSP, based on the Variational Autoencoder (VAE) paradigm, excels in both crystal structure reconstruction and ab initio generation tasks. Experimental results show exceptional performance across evaluation metrics. Compared to previous work\cite{xie2021crystal}, we have introduced innovations and conducted verifications in crystal structure representation and the encoder network.}
\label{introduce_fig}
\end{center}
\vskip -0.2in
\end{figure*}

CSP methods based on generative models have progressively evolved from utilizing composition information to directly constructing complex crystal structures. Early works, such as MatGAN\cite{dan2020generative} and Binded-VAE\cite{oubari:hal-04464893}, rely exclusively on crystal structure composition information for training, capturing implicit rules of chemical composition. However, these models serve primarily as reference models for structure generation and have a limited capacity to produce diverse structures. More advanced methods, including Composition-Conditioned Crystal GAN\cite{kim2020generative} and CCDCGAN\cite{long2021constrained}, have made significant progress in feature representation, but they still constrain the crystal structure space and are limited to specific compound groups. In contrast, works for complex systems, such as CDVAE\cite{xie2021crystal}, DiffCSP\cite{jiao2023crystal}, UniMat\cite{yang2024scalable}, and MatterGen\cite{zeni2025generative}, have achieved crystal structure generation without being restricted by representation space, aligning with the primary research goals in materials discovery. Therefore, optimizing existing model frameworks and exploring new approaches to continuously enhance both the efficiency and quality of structure generation represent key directions for advancing the application of artificial intelligence in materials science.

The development of CSP methods highlights that the accuracy of algorithms and models for crystal structure characterization is a critical factor influencing the stability and precision of generated structures. This accuracy remains a significant challenge in generative models for CSP. In particular, capturing the periodicity, symmetry, and other fundamental characteristics of crystal structures continues to be a major obstacle in CSP methodologies. To address these challenges, we propose a Transformer-Enhanced Variational Autoencoder (TransVAE-CSP) for crystal structure prediction, as shown in Fig. \ref{introduce_fig}. This model facilitates the accurate reconstruction and generation of crystal structures. The primary contributions of this work are as follows:
\begin{itemize}
    \item We developed an adaptive distance expansion method to characterize crystal structures and introduced an innovative hybrid function method to optimize a single distance expansion function. Our verification showed that this hybrid strategy can yield optimal results under different crystal sample environments;
    \item We proposed a Transformer network based on an equivariant dot product attention mechanism as encoder for Variational Autoencoder (VAE) model, enhancing its ability to learn the characteristics of crystal samples;
    \item Our work was evaluated using the carbon\_24, perov\_5 and mp\_20 datasets. The results demonstrated significant advantages in structure reconstruction and generation tasks under various modeling metrics. This research offers a powerful tool for crystal structure design and optimization.
\end{itemize}

\section{Related Work}
In this section, we will introduce the primary challenges encountered by CSP methods (Section \ref{profor}), commonly used crystal structure characterization techniques (Section \ref{cryrep}), encoding network models (Section \ref{equnet}) and crystal generation models (Section \ref{crygenmodel}).

\subsection{Problem Formulation}
\label{profor}
A crystal structure consists of atoms interacting with each other, where each atom is characterized by its element type and coordinates. The goal of learning its representation is to train an encoder that accurately maps the crystal structure into a representation within a sampling space. To preserve the properties of the crystal structure after transformations, the structure must maintain translation, rotation, inversion, and permutation invariance\cite{keating1966effect, lin2022general}. Among these, translation, rotation, and inversion operations form E(3) symmetry, while translation and rotation operations form SE(3) symmetry. However, current models often struggle to effectively capture key features of crystal structures like symmetry and periodicity, particularly global symmetry, which can result in generated crystal structures failing to retain correct physical properties. Effectively incorporating these symmetries into material structure representation and neural networks has emerged as a significant challenge in crystal structure prediction. This issue also represents a classic problem in structural encoding.

\subsection{Crystal Representation}
\label{cryrep}

The crystal structure representation methods form the foundation for studying the relationship between crystal properties and structures. Previous representation techniques, such as those based on simple matrices\cite{oubari:hal-04464893, dan2020generative, kim2020generative, long2021constrained}, are inadequate for representing complex systems. The advent of graph representation methods has significantly improved the ability of models to learn crystal properties\cite{xie2018crystal, schutt2018schnet, chen2019graph, chen2022universal, yuan2024tripartite}. Notably, the incorporation of information regarding three-body interactions has further improved the predictive power of these models.

\subsection{Equivariant Network}
\label{equnet}
To facilitate the representation of crystal structure symmetries in generative network models, e3nn \cite{mario_geiger_2022_6459381}, based on previous work\cite{Thomas2018TensorFN, weiler20183d}, is specifically designed to handle the symmetry of E(3). E3nn simplifies the process of constructing and training neural networks with these characteristics.

Equivariant networks\cite{kondor2018clebsch,unke2021se, tholke2022equivariant, brandstetter2022geometric} utilize geometric functions constructed from spherical harmonics and irreducible features to implement 3D rotation and translation equivariance, as proposed in Tensor Field Networks (TFN)\cite{Thomas2018TensorFN}. The SE(3) Transformer\cite{fuchs2020se} employs equivariant dot product (DP) attention\cite{vaswani2017attention} with linear messages. Equiformer \cite{liao2023equiformer} integrates MLP attention with non-linear messages and various types of support vectors, enhancing the model's expressiveness.

\subsection{Crystal Generative Model}
\label{crygenmodel}
The generative model paradigm has been extensively utilized to generate material structures, including VAEs\cite{oubari:hal-04464893, noh2019inverse, hoffmann2019data, Ren2020InverseDO, court20203, xie2021crystal}, GANs\cite{nouira2018crystalgan, dan2020generative, kim2020generative, long2021constrained}, and diffusion models\cite{jiao2023crystal, yang2024scalable, zeni2025generative}. Most of these studies focus on binary compounds\cite{noh2019inverse, long2021constrained}, ternary compounds\cite{nouira2018crystalgan, kim2020generative}, or simpler materials within the cubic system\cite{hoffmann2019data, court20203}. Xie et al.\cite{xie2021crystal} incorporated the diffusion concept into the decoder component of the VAE, addressing the challenge of predicting material coordinates, while also retaining the crystal reconstruction capabilities of the VAE. The diffusion model\cite{jiao2023crystal, yang2024scalable, zeni2025generative} facilitates applications in more complex systems, such as the Materials Project dataset\cite{jain2013commentary}, and is capable of performing conditional generation tasks.

\section{Methods}
\begin{figure*}[ht]
\vskip 0.2in
\begin{center}
\centerline{\includegraphics[width=(\columnwidth*2)]{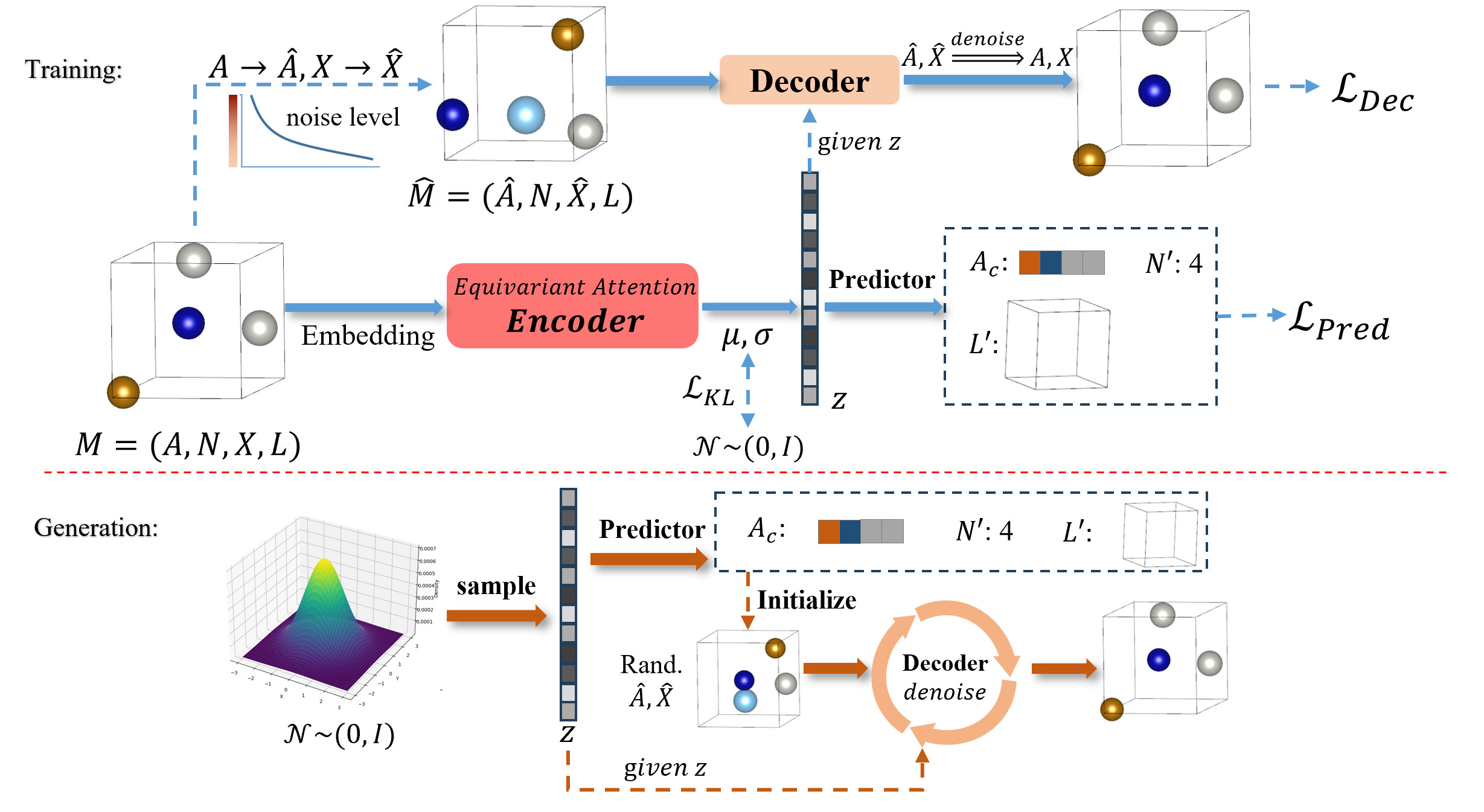}}
\caption{Overview of TransVAE-CSP. Training VAE model: Given $M=(A,N,X,L)$, it captures feature information through the \textbf{Embedding} layer and the equivariant attention encoder to obtain the latent space variable $z$, which is utilized as a condition to guide the output of both the \textbf{Predictor} and the denoising \textbf{Decoder}. The loss function comprises three parts:$\mathcal{L}_{Pred}, \mathcal{L}_{Dec}, \mathcal{L}_{KL}$. Generation: The variable $z$ is sampled from a multi-dimensional standard normal distribution $\mathcal{N}\sim(0,I)$ and ab initio generation is achieved by the \textbf{Predictor} and the denoising \textbf{Decoder} to produce a crystal structure that aligns with the feature space of the training samples. Note: Our work optimizes the algorithm based on CDVAE, thereby the framework refers to Xie et al.\cite{xie2021crystal}.}
\label{overview_fig}
\end{center}
\vskip -0.2in
\end{figure*}

In this section, we will provide a comprehensive introduction to the TransVAE-CSP that we developed. As illustrated in Fig. \ref{overview_fig}, the model consists of two primary components: training and generation. It uses well-established and highly accurate CDVAE. Building upon this work, we have implemented a new encoder network to train a high-quality learning model.

During the training process, TransVAE-CSP focuses primarily on optimization of the crystal structure representation and the encoder neural network. In Sections \ref{ade} and \ref{equdpan}, we will elucidate the design concepts and work principles of our proposed adaptive distance expansion method for crystal structure representation, as well as the Transformer network encoder that utilizes the equivariant dot product attention mechanism. We provide a detailed explanation of how to capture the periodicity and symmetry characteristics of crystal structures, thereby enabling effective encoding of the crystal structure. Finally, Section \ref{ooa} will present a comprehensive overview of the module design and workflow of TransVAE-CSP.

\subsection{Adaptive Distance Expansion}
\label{ade}
The distance between atoms in a crystal structure directly influences its stability and physical properties. Employing mathematical methods or functions to describe the relationships between atomic distances is a common approach in the analysis of crystal structures within the field of machine learning. This includes techniques such as distance matrices, Fourier series expansions, and radial basis functions. In materials design, particularly in high-throughput screening and inverse material design, predicting material performance requires an efficient computational model. The radial basis function is particularly effective at capturing the complexities of nonlinear relationships, allowing it to flexibly manage intricate structural interactions. Furthermore, it can be utilized to establish a mapping between material structure and physical properties, facilitating material screening and optimization\cite{gasteiger_dimenetpp_2020, gasteiger2021gemnet, choudhary2021atomistic, chen2022universal, liao2023equiformer}.

The distance expansion adopts the Radial Basis Function (RBF) method\cite{buhmann2000radial}, which is expressed as $\phi(x)=\varphi(\|x-c\|)$. $x$ is the input vector, $c$ is the center point, $\|x-c\|$ represents the Euclidean distance from the input $x$ to the center point $c$. $\varphi$ is a function, usually selected as some smooth function, such as Gaussian function\cite{fornberg2011stable}, Bessel function\cite{javaran2011first}, etc.

\textbf{Gaussian RBF} Its function value decreases rapidly as the distance between the input $x$ and the center point $c$ increases. The Gaussian RBF exhibits characteristics of locality, smoothness, and nonlinear mapping. It transforms distances into a high-dimensional feature space, allowing the original low-dimensional input data to be mapped to a higher-dimensional feature space, thereby facilitating the embedding of distance features. For detailed information on Gaussian RBF, please refer to Appendix \ref{appendix:gaussian_rbf}

\textbf{Bessel RBF} Because Bessel RBF exhibits distinct oscillation and periodic characteristics compared to other radial basis functions, such as the Gaussian RBF, so it offer unique advantages in specific periodic data processing tasks. The function is typically constructed using Bessel function.For detailed information on Gaussian RBF, please refer to Appendix \ref{appendix:bessel_rbf}

\textbf{Hybrid RBF} To effectively capture the intricate relationships within crystal structures and enhance the model's capacity to accommodate these complexities, we propose a hybrid radial basis function method that integrates the Gaussian radial basis function with the Bessel radial basis function. This approach employs a function fusion algorithm based on weighted concatenation. To address the issue of differing orders of magnitude in the eigenvalues of the two functions, we implement a weight scaling algorithm. The formula for the hybrid radial basis function is expressed as: $H(x) = \Phi(x)\bigoplus k\cdot\Psi(x)$, where $k$ denotes the weight, $\Phi$ represents the Bessel RBF, $\bigoplus$signifies the concatenation operation, and $\Psi$ denotes the Gaussian RBF.

\textbf{Overall}, when training the model, we adaptively employ various radial basis functions for distance expansion across different data sets by adjusting the coefficients. We then compare the training results and select the radial basis function that performs best for each specific data set.

\begin{figure}[htbp]
  \centering
  \includegraphics[width=(\columnwidth*4/5)]{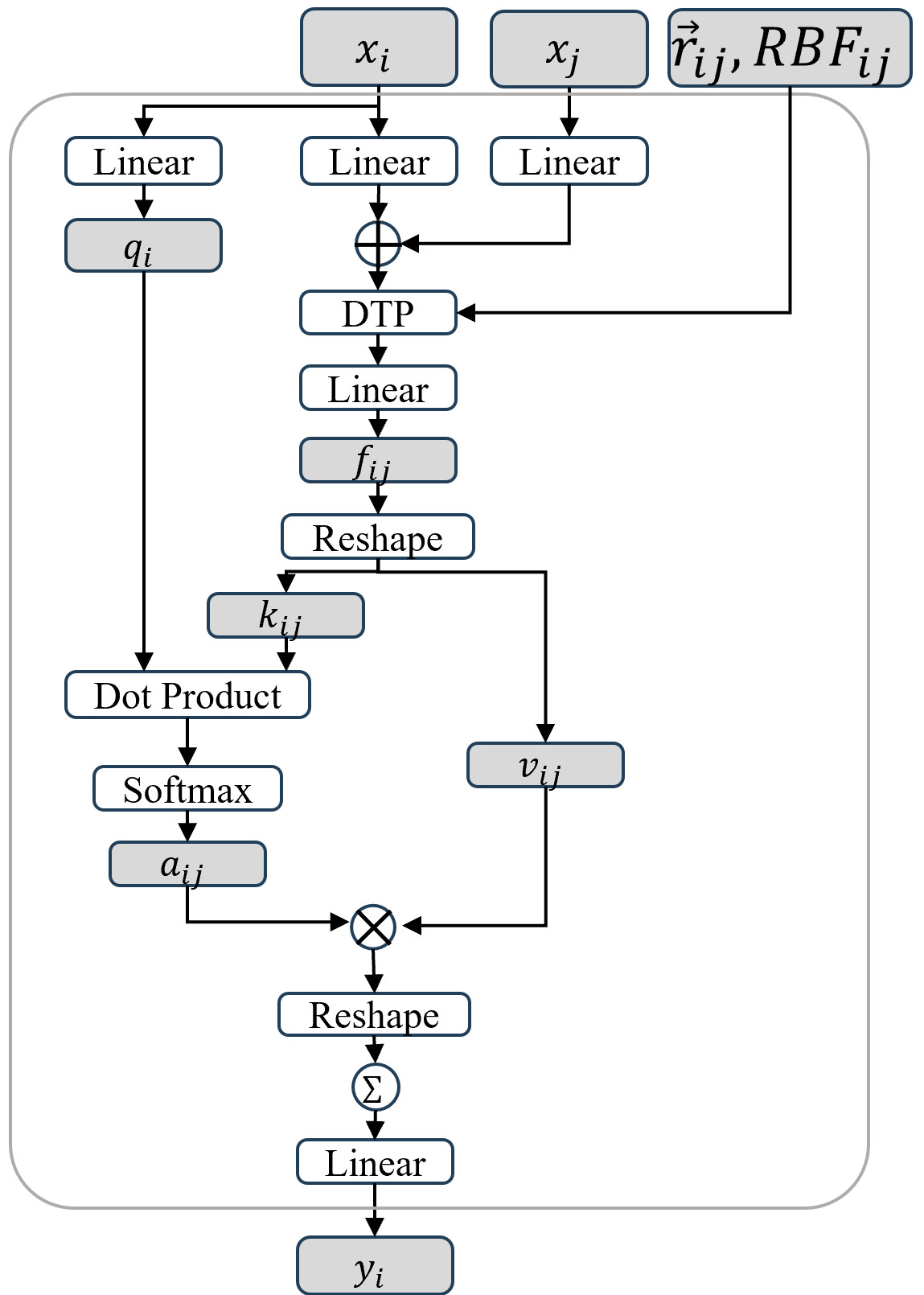}
  \caption{The technique of equivariant dot product attention network. This is the core network of Transformer. “DTP” stands for depth-wise tensor product, $\bigoplus$ denotes addition, $\bigotimes$ denotes multiplication.$\sum$ denotes scatter operation.}
  \label{fig:equnet}
\end{figure}
\subsection{Equivariant Dot Product Attention Network}
\label{equdpan}
The encoder utilizes an adapted SE(3)-Transformer network \cite{fuchs2020se}, with the core structure being the equivariant dot product attention network \cite{liao2023equiformer}, as shown in Fig. \ref{fig:equnet}. Node \(i\) serves as the center, and the purpose of this network is to analyze the neighboring environmental information of the central node to obtain its higher-order features. \(X_i\) outputs the query \(q_i\) through a linear layer \cite{liao2023equiformer}. The features of \(X_i\) and its neighbor node \(X_j\) are combined after passing through linear layers to obtain the initial message:$x_{ij} = \text{Linear}_{\text{dist}}(x_{i}) + \text{Linear}_{\text{src}}(x_{j})$. Then, \(x_{ij}\) and its distance expansion \(\vec{r}_{ij}\) are processed via the DTP network \citep{howard2017mobilenets}:
\begin{equation}
\label{dtp}
f_{ij} = \text{Linear}\left( 
  x_{ij} \otimes_{\omega\left(\left\Vert \vec{r}_{ij} \right\Vert\right)}^{\text{DTP}} 
  \Bigl[ \text{SH}(\vec{r}_{ij}),\, \text{RBF}_{ij} \Bigr] 
\right)
\end{equation}
SH is the spherical harmonic embedding, \(\omega(\| \vec{r}_{ij} \|)\) is the weight of the Tensor Product. \(F_{ij}\) represents the fused features of the target node and the source node, and is used to derive the attention weights and messages. We split \(f_{ij}\) into two incompatible features, key \(k_{ij}\) and value \(v_{ij}\). Then, we perform a scaled dot product between \(q_i\) and \(k_{ij}\) \cite{NIPS2017_3f5ee243} to obtain the attention weights. Finally, the weights and messages are multiplied, and the higher-order feature messages of all target nodes are obtained through Reshape and aggregation functions, as well as linear layers:
\begin{equation}
y_i = \text{Linear}\left( \sum_{\text{scatter}} a_{ij} \otimes v_{ij} \right)
\end{equation}
The scatter operation performs a summation aggregation on the messages calculated for the same target atom and all its neighbors according to the index, in order to obtain the environmental feature message for each atom.

\subsection{Overall of Architecture}
\label{ooa}
Figure \ref{overview_fig} shows the framework of TransVAE-CSP. Building on the work of CDVAE \cite{xie2021crystal}, our research focuses on optimizing crystal structure representation and the encoder network.

\textbf{Representation} Any crystal structure is an infinite repetition of a unit cell in 3D space. A unit cell can be represented as follows, where \( A \) represents the type of atoms, \( N \) represents the number of atoms, \( X \) represents the atomic coordinates, and \( L \) represents the lattice parameters.If the matrix \( N \) contains elements in \( \mathbb{R} \), let \( A = (a_{0}, a_{1}, \ldots, a_{n-1}) \in \mathbb{E}^{N} \) represent the set of all elements in the matrix. Let \( X = (x_{0}, x_{1}, \ldots, x_{m-1}) \in \mathbb{R}^{N \times 3} \) and \( L = (a, b, c, d, e) \in \mathbb{R}^{6} \).

\begin{figure}[htbp]
  \centering
  \includegraphics[width=\columnwidth*4/5]{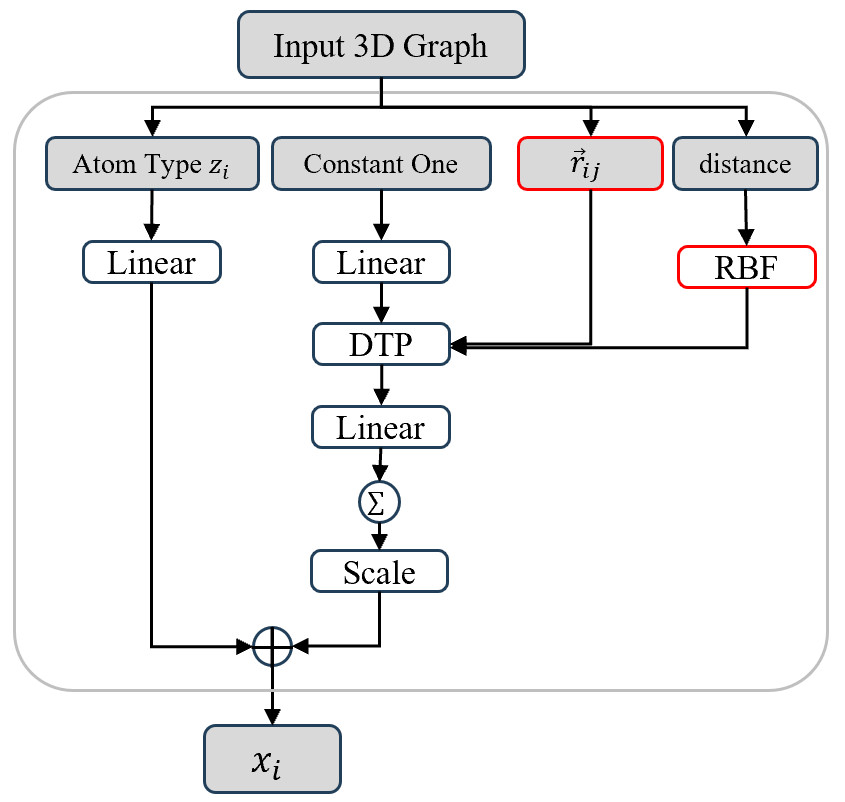}
  \caption{Embedding block. Compared to pervious work\cite{liao2023equiformer}, our adaptation work is highlighted in red, and \textbf{RBF} refers to the feature vector obtained after the radial basis function is expanded based on the distance.}
  \label{fig:embedding}
\end{figure}

\textbf{Embedding} This module is adapted from\cite{liao2023equiformer}, as shown in Fig. \ref{fig:embedding}, and consists of Atom Embedding and Edge Degree Embedding. Atom Embedding uses a linear layer to one-hot encode the atom type. Edge Degree Embedding first expands the RBF features (marked in red) of the constant vector, interaction information, and atom pair distance. It then encodes the local geometric message through two linear layers and an intermediate DTP layer, followed by aggregation to encode the message. The form of the DTP layer is the same as in formula \ref{dtp}. Finally, the aggregated features are scaled by dividing by the square root of the average degree in the training set (the calculation method is provided in Appendix \ref{appendix:gvfd}) to ensure that the standard deviation of the aggregated features is close to 1. After adding the Atom Embedding and Edge Degree Embedding together, the feature embedding of the final node is obtained.

\textbf{Encoder} The encoder employs a transformer network based on the equivariant dot product attention mechanism (refer to Section \ref{equdpan}). The final output is generated through aggregation operations to derive the intermediate variables of the VAE, which are subsequently used as inputs to the mean and linear layers.

\textbf{Decoder} The decoder consists of two parts: multiple independent MLP networks that predict the atomic type components \( A_c \), the number of atoms \( N \), and the lattice parameters \( L \). The prediction of atomic coordinates and types is based on the diffusion concept, incorporating random noise and employing the GemNet-T network\cite{gasteiger2021gemnet} to predict the atomic type and coordinate noise of the input structure.

\textbf{Training and Generation} In the training phase, the encoder generates the intermediate variable \( z \), which is used as the condition for the denoising encoder and as input for the predictor. The loss function comprises the aggregation loss from the predictor, the loss from the denoising network, and the KL divergence loss (see Appendix \ref{appendix:loss_fuc} for details). In the generation phase, the intermediate variable \( Z \) is sampled from the standard normal distribution, and the structural components \( A_c \), the number of atoms \( N \), and the lattice parameters \( L \) are generated through the predictor. The coordinate information and atomic types are randomly generated as the initial structure. Given \( z \) as the condition, annealed Langevin dynamics\cite{song2019generative} is performed through the denoising decoder to continuously update the atomic types and coordinates.

\section{Results}

We evaluated the effectiveness of the fusion model in various tasks related to generating crystal structures and used the test standards from Xie et al. \cite{xie2021crystal} to compare the results. We assessed our work in \ref{cryrecon} crystal structure reconstruction and \ref{crystrgen} crystal structure generation. Additionally, we demonstrate in \ref{comprbf} that the radial basis function exhibits varying adaptability across different datasets. The datasets employed in the experiment consist of three high-quality datasets of differing complexities. Perov-5 \cite{castelli2012new, castelli2012computational} focuses on perovskite materials and contains 18,928 perovskite materials with similar structures but different elemental compositions, including 56 elements, with each structure containing 5 atoms per unit cell. Carbon-24 \cite{pickard2020airss} focuses on carbon-based materials and contains 10,153 carbon materials, which consist solely of carbon elements, with each unit cell containing between 6 and 24 atoms. MP-20 selects 45,231 stable inorganic materials from Materials Project \cite{jain2013commentary}, which includes most experimentally generated materials with up to 20 atoms per unit cell. We applied a 60-20-20 split in accordance with Xie et al. \cite{xie2021crystal}.

\begin{table*}[t]
\caption{Reconstruction Performance of Different Models.}
\label{tab:recon_perform}
\vskip 0.15in
\begin{center}
\begin{small}
\begin{sc}
\begin{tabular}{lcccccc}
        \toprule
        \textbf{Model} & \multicolumn{2}{c}{\textbf{Perov\_5}} & \multicolumn{2}{c}{\textbf{Carbon\_24}} & \multicolumn{2}{c}{\textbf{MP\_20}} \\
        & \textbf{Match Rate} & \textbf{RMSE} & \textbf{Match Rate} & \textbf{RMSE} & \textbf{Match Rate} & \textbf{RMSE} \\
        \midrule
        FTCP & \textbf{99.34} & 0.0259 & 62.28 & 0.2563 & 69.89 & 0.1593 \\
        Cond-DFC-VAE & 51.65 & 0.0217 & - &  - & - & - \\
        CDVAE & 97.52 & 0.0156 & 55.22 & 0.1251 & 45.43 & \textbf{0.0356} \\
        TransVAE-CSP & 98.19 & \textbf{0.0092} & \textbf{80.75} & \textbf{0.0938} & \textbf{71.14} & 0.0377 \\
    \bottomrule
    \end{tabular}
\end{sc}
\end{small}
\end{center}
\vskip -0.1in
\end{table*}

\begin{table*}[t]
\caption{Real Performance of TransVAE-CSP reconstruction structures. }
\label{recon_compar}
\vskip 0in
\begin{center}
\begin{small}
\begin{sc}
\begin{tabular}{ccccccc}
\toprule
 & \multicolumn{2}{c}{Perov\_5} & \multicolumn{2}{c}{Carbon\_24} & \multicolumn{2}{c}{MP\_20} \\
\midrule
Ground Truth & \includegraphics[width=2cm]{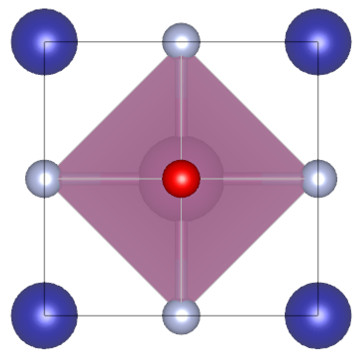} & \includegraphics[width=2cm]{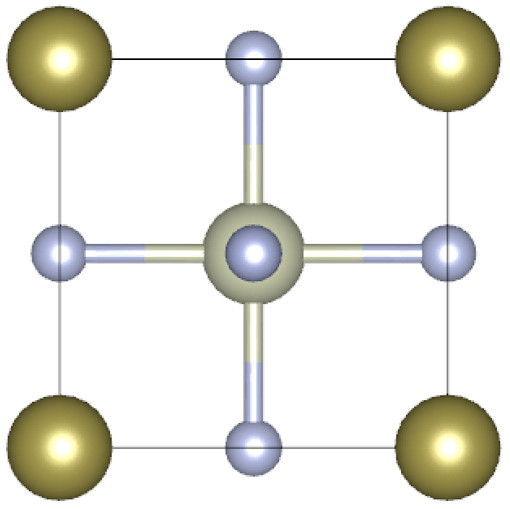} & \includegraphics[width=2cm]{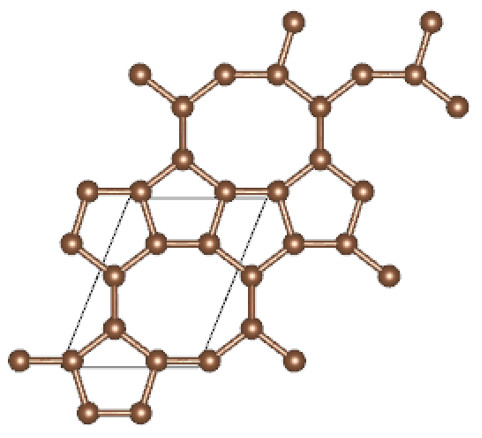} & \includegraphics[width=2cm]{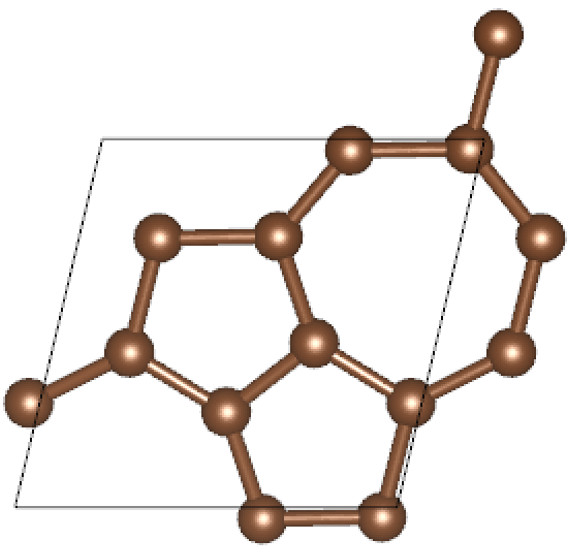} & \includegraphics[width=2cm]{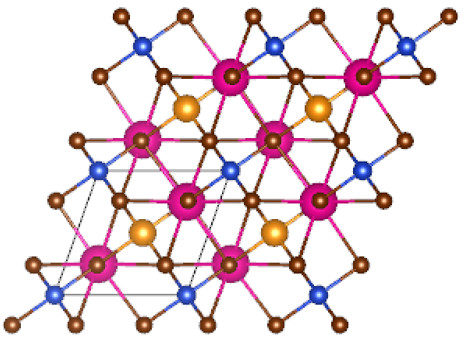} & \includegraphics[width=2cm]{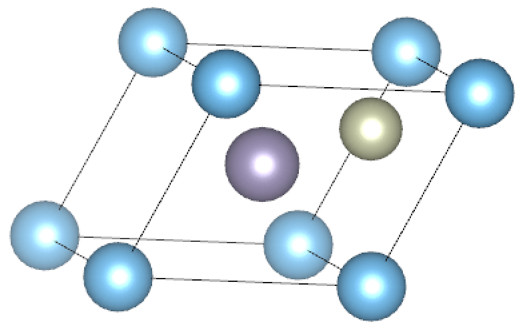} \\
Prediction   & \includegraphics[width=2cm]{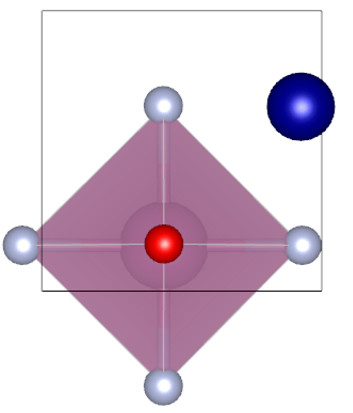} & \includegraphics[width=2cm]{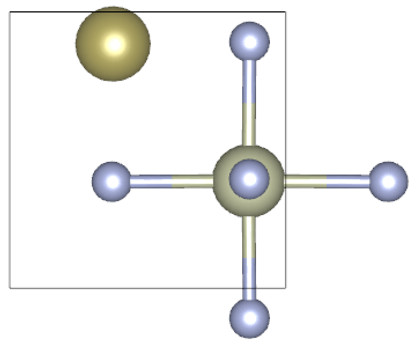} & \includegraphics[width=2cm]{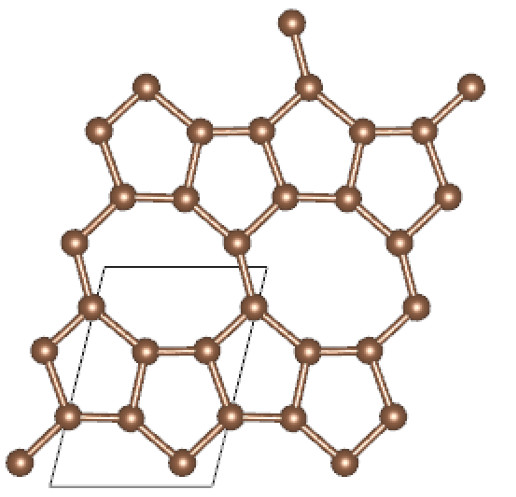} & \includegraphics[width=2cm]{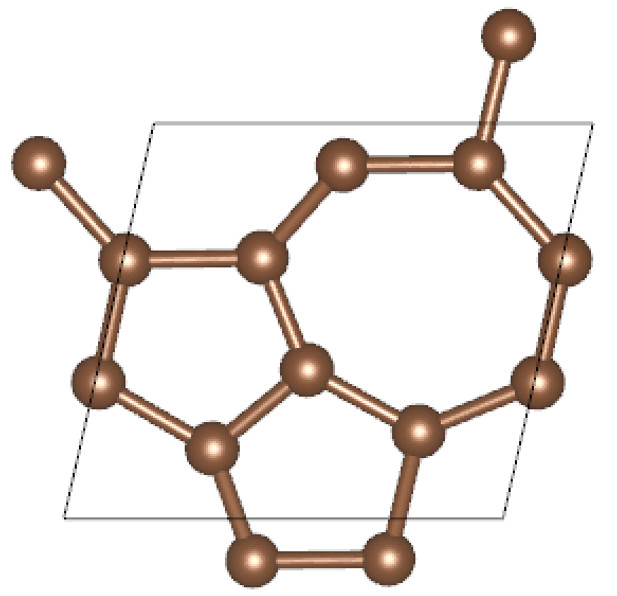} & \includegraphics[width=2cm]{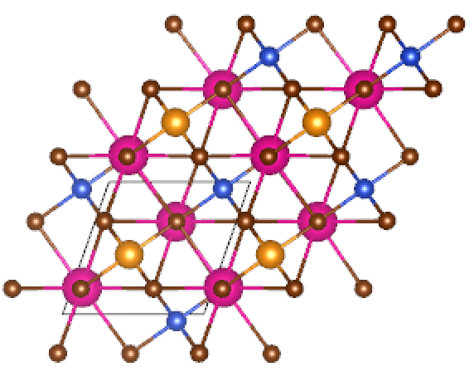} & \includegraphics[width=2cm]{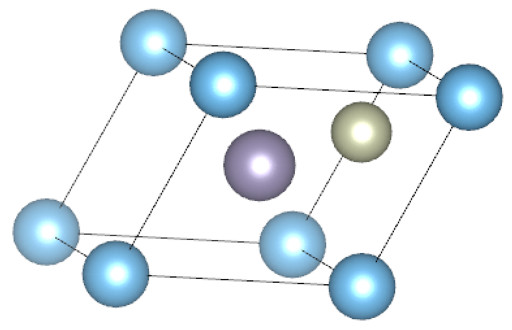}\\

\bottomrule
\end{tabular}
\end{sc}
\end{small}
\end{center}
\vskip -0.1in
\end{table*}

\subsection{Crystal Structure Reconstruction}
\label{cryrecon}
\subsubsection{Baseline}
In the crystal reconstruction task, we compared our model with three baselines from \cite{xie2021crystal}: FTCP \cite{Ren2022}, Cond-DFC-VAE \cite{court20203}, and CDVAE \cite{xie2021crystal}. CDVAE encodes the atoms, atomic pair distances, and angles of three atoms in the crystal structure. It predicts the composition, lattice, and number of atoms from the latent variable space \( z \), initializes a crystal structure with random atomic types and coordinates, uses the Langevin dynamics method to denoise the atomic types and coordinates, and finally generates a new structure.

\subsubsection{Metrics}
We follow the common practice \cite{xie2021crystal,jiao2023crystal} of predicting crystal structures by testing the latent intermediate variables corresponding to all structures in the test set. We use the \texttt{StructureMatcher} class from \texttt{Pymatgen}, setting the thresholds to \texttt{stol}=0.5, \texttt{angle\_tol}=10, and \texttt{ltol}=0.3 to measure whether the real structure matches the reconstructed structure, thus calculating the match rate to evaluate the reconstruction performance. The RMSE is calculated as the average over the successfully matched reconstructed structures and real structures, normalized by \(\sqrt[3]{V/N}\), where \(V\) is the volume of the structure's unit cell and \(N\) is the number of atoms in the unit cell.

\begin{table*}[th]
\caption{Generation Performace}
\label{generation_performace}
\vskip 0.15in
\begin{center}
\begin{small}
\begin{sc}
\begin{tabular}{ccccccccc}
\toprule
 \multirow{2}*{\textbf{Dataset}} & \multirow{2}*{\textbf{Model}} & \multicolumn{2}{c}{\textbf{Validity}(\%) $\uparrow$} & \multicolumn{2}{c}{\textbf{Coverage}(\%) $\uparrow$} & \multicolumn{3}{c}{\textbf{Property statistics} $\downarrow$} \\
 ~ & ~ & Stru. & Comp. & Cov-R & Cov-P & $d_{density}$ & $d_E$ & $d_{elem}$ \\
\midrule
\multirow{7}*{perov\_5} & FTCP & 0.24 & 54.24 & 0.0 & 0.0 & 10.27& 156 & 0.6297 \\
& Cond-DFC-VAE & 73.6&	82.95&73.92&10.13&	2.268&	4.111	&0.8373 \\
& G-SchNet& 99.92&	98.79&	0.18&	0.23&	1.625&	4.746&	0.0368 \\
&P-G-SchNet &79.63&	99.13&	0.37&	0.25&	0.2755&	1.388&	0.4552\\
&CDVAE &100	&98.59	&99.45&	\textbf{98.46}&	0.1258&	0.0264&	0.0628 \\
& DiffCSP&100&	\textbf{98.85}&	\textbf{99.74}&	98.27&	0.111&	 \textbf{0.0263}&	\textbf{0.0128} \\
& TransVAE-CSP &\textbf{100}	&98.82&	99.52&	98.33&	\textbf{0.1082}&	0.0311&	0.0992 \\
\hline
\multirow{6}*{carbon\_24} & FTCP &0.08&	-&	0.0&	0.0&	5.206&	19.05&	- \\
& G-SchNet&99.94&	-&	0.0&	0.0&	0.9427&	1.32&	- \\
&P-G-SchNet &48.39&	-&	0.0&	0.0&	1.533&	134.7&	- \\
&CDVAE & \textbf{100} &	- &	99.8&	83.08&	0.1407&	0.285&	- \\
& DiffCSP&\textbf{100} &	-	& 99.9 &	\textbf{97.27}&	\textbf{0.0805}&	\textbf{0.082}&	- \\
& TransVAE-CSP &99.9& - &\textbf{100}	&78.62	&0.1636	&1.744	&- \\
\hline
\multirow{6}*{mp\_20} & FTCP &1.55 &	48.37&	4.72&	0.09	&23.71	&160.9&	0.7363 \\
& G-SchNet&99.65&	75.96&	38.33&	99.57&	3.034&	42.09&	0.6411 \\
&P-G-SchNet &77.51&	76.4&	41.93&	99.74&	4.04&	2.448&	0.6234 \\
&CDVAE & 100&	86.7&	99.15&	99.49&	0.6875&	0.2778&	1.432\\
& DiffCSP&100&	83.25&	\textbf{99.71}&	\textbf{99.76}&	0.3502&	0.1247&	\textbf{0.3398} \\
& TransVAE-CSP & \textbf{100}&	\textbf{87.97}&	99.67&	99.23&	\textbf{0.1694}&\textbf{0.1182}&	0.7348 \\
\bottomrule
\end{tabular}
\end{sc}
\end{small}
\end{center}
\vskip -0.1in
\end{table*}
\subsubsection{Results}
The indicators of the reconstructed structure are presented in Table \ref{tab:recon_perform}. Unlike previous studies, we adaptively employ different radial basis functions (RBFs) for distance expansion across various data sets by adjusting coefficients during model training. We compare the training outcomes and automatically select the radial basis function that demonstrates the best performance for each data set. The optimal results are achieved using Gaussian, Hybrid, and Bessel RBFs for MP\_20, Carbon\_24, and Perov\_5, respectively. This selection is influenced by the structural complexity of the data set, the long-range interactions between atoms, and the diversity of samples, as well as the adaptability of RBFs within the atomic pair distance distribution space of different data sets. In terms of indicator performance, our model exhibits a slightly lower Match Rate than FTCP for Perov\_5 and a marginally higher RMSE than CDVAE for MP\_20. Other indicators surpass the baseline models to varying extents. For detailed information on model training, please refer to Appendix \ref{appendix:train_para}.

The comparison between the reconstructed structure and the real structure is shown in Table \ref{recon_compar}. The observation direction of each pair of structures remains the same. Considering that our model maintains the invariance of 3D space under translation, rotation, and periodicity, the visualization results of atoms at corresponding symmetric points in the reconstructed and real structures may differ due to their different origin positions. However, by applying lattice symmetry operations (such as translation), the alignment of atomic positions within the unit cell can be achieved.

\subsection{Crystal Structure Generation}
\label{crystrgen}
\subsubsection{Baseline}
For structure generation, we adopt ab initio crystal generation. We compare our model with six baselines \cite{jiao2023crystal}: FTCP \cite{Ren2022}, Cond-DFC-VAE \cite{court20203}, G-SchNet, P-G-SchNet \cite{gebauer2019symmetry}, CDVAE \cite{xie2021crystal}, and DiffCSP \cite{jiao2023crystal}. DiffCSP requires the input of specified atom types and the number of atoms to complete the generation task. For consistency, we sample the number of atoms from the pre-computed distribution in the training set \cite{hoogeboom2022equivariant}, allowing DiffCSP to achieve the same ab initio crystal generation as the other works.

\subsubsection{Metrics}
Similar to \cite{xie2021crystal}, we employ three metrics to evaluate generation performance: Validity (including Structure Validity and Composition Validity), Coverage (comprising Coverage Recall and Coverage Precision), and Property Statistics. These metrics assess the validity of the predicted crystal, the similarity between the test set and the generated samples, and the distribution distance of property calculations related to density, formation energy, and the number of elements. For detailed information on the metrics, please refer to Appendix \ref{appendix:metric}.

\subsubsection{Results}
The results of the indicators are shown in Table \ref{generation_performace}. Our model performs well across all these indicators. On the perov\_5 dataset, although our model does not perform optimally in terms of coverage, COV-R and COV-P are in the middle range of the corresponding indicator values of the two optimal baseline models, and our model achieves the best performance in terms of density distribution distance. On the carbon\_24 dataset, our model achieves 100\% in the COV-R indicator, but COV-P does not perform as well. On the mp\_20 dataset, the composition validity, density distribution distance, and energy distribution distance indicators exceed those of all baselines, and the other indicators also achieve excellent performance.

The results show that our work performs best on the mp\_20 dataset, which may be related to the authenticity of the structures in the mp\_20 dataset. In the perov\_5 and carbon\_24 datasets, most structures are thermodynamically unstable, which is also a factor affecting the model's performance.

\subsection{Comparison of RBF Strategies}
\label{comprbf}
We explored the impact of radial basis function expansion methods for encoding interatomic distances in crystal structure characterization on the model. Here, we conducted comparative tests on radial basis functions based on Gaussian functions, Bessel functions, and hybrid functions of the two on the mp\_20, carbon\_24, and perov\_5 datasets. The test indicator is the convergence of the loss value over 200 iterations on the training set and the validation set during training. The results are shown in Appendix \ref{appendix:rbf}. On the mp\_20, carbon\_24, and perov\_5 datasets, the model optimization effects of the Gaussian, hybrid, and Bessel radial basis function distance expansion methods are the best.

\section{Conclusion and Outlook}
\label{conclusion}
In this work, we propose a crystal structure generation method based on the Transformer-Enhanced Variational Autoencoder for Crystal Structure Prediction (TransVAE-CSP). Through experiments, we demonstrate that the equivariant dot-product attention Transformer network outperforms traditional Graph Convolutional Networks (GCNs) in understanding the chemical and physical properties of crystal structures. We innovatively introduce an adaptive distance expansion method, and the experimental results further validate that this approach achieves the desired objectives, providing a new optimization perspective for crystal structure representation. Moreover, compared to diffusion models, the Variational Autoencoder (VAE) shows unique advantages in structure reconstruction and distribution consistency verification, highlighting its potential in crystal structure generation.

Future work will optimize the VAE-based crystal structure generation model to match the diffusion models, while enhancing the framework for efficient prediction of structures with specified elemental compositions. This advancement will drive crystal structure prediction technologies and enable innovative design of high-performance materials.


\section*{Impact Statement}
This paper presents work whose goal is to advance the field of 
Machine Learning. There are many potential societal consequences 
of our work, none which we feel must be specifically highlighted here.

\nocite{langley00}

\bibliography{main-reference}
\bibliographystyle{icml2025}

\newpage
\appendix
\section{Radial Basis Function (RBF)}
\subsection{Gaussian RBF}
\label{appendix:gaussian_rbf}
The gaussian RBF formula is followed:
\begin{equation}
    \label{gaussian_base_equation}
    \phi_i(x)=exp(-\frac{\|x-c_i\|^2}{2\sigma_i^2})
\end{equation}
where $c_i$ is the center point, $\sigma_i$ is the breadth of basis functions. The Gaussian RBF expression for mapping distances into N-dimensional feature spaces is represented by Formula \ref{gaussian_expand_equation}, which requires initializing $N$ parameters $ci$ and $\sigma_i$.

\begin{equation}
    \label{gaussian_expand_equation}
    \Phi(x)=[\phi_1(x), \phi_1(x), \phi_2(x), ..., \phi_N(x)]
\end{equation}
\subsection{Bessel RBF}
\label{appendix:bessel_rbf}
Bessel RBF is typically constructed using Bessel function. The formula is followed as: $\phi(x)=J_v(\|x-c\|)$, where J(r) represents the v-th order Bessel function, where v is typically a non-negative real number. The variable r is defined as the Euclidean distance between the input vector x and the center point c, expressed as r = ‖x - c‖. The distance is transformed into a high-dimensional feature space using the following formulation.
\begin{equation}
\begin{split}
    \Phi(x)=[
    & J_v(\|x-c_1\|)\cos{(\omega_1\|x-c_1\|)}, \\
    & J_v(\|x-c_2\|)\cos{(\omega_2\|x-c_2\|)}, \\
    & ..., \\
    & J_v(\|x-c_n\|)\cos{(\omega_n\|x-c_n\|)}]
\end{split}
\end{equation}
Among the formula, $\omega$ denotes the frequency parameter, which is utilized to regulate the oscillation frequency of the function, while $\cos$ represents the cosine function.

\section{Experiment Details}
\subsection{Global Variables for Datasets}
\label{appendix:gvfd}
The encoder network must provide several statistics from the sample dataset: the maximum number of nodes, the average number of nodes, and the average coordination number. These statistics are essential for performing more accurate scaling operations across different sample spaces within the network. The maximum number of nodes refers to the highest number of atoms present in the unit cell across all crystal structures in the statistical dataset. The average number of nodes is determined by calculating the total number of nodes for all structures in the sample dataset, expressed mathematically as $\frac{1}{N}\sum_{i=1}^N NUM_{node}$, where $NUM_{node}$ denotes the number node of a structure in a unit cell, $N$ denotes the total number of all structures. The coordination number of each atom is computed using the \texttt{CrystalNN} class from the \texttt{Pymatgen} library, which is a tool designed for analyzing atomic neighbors in crystalline materials. The method `get\_cn()` is a crucial function of the \texttt{CrystalNN} class, utilized to calculate the coordination number (CN, also called \textbf{degree}) of a specific atom. The coordination number represents the number of directly connected neighboring atoms surrounding a given atom and is typically used to describe the local environment of that atom within the crystal structure. The formula for calculating the average coordination number is
\begin{equation}
    AVG_{degree}=\frac{1}{N}\sum_{i=1}^N (\frac{1}{M}\sum_{j=1}^M CN_j),
\end{equation}
where $N$ denotes the total number of all structures, $M$ denotes the total number of node in a structure, $CN_j$ denotes the coordination number of a node in a structure.
\subsection{Training Parameters}
\label{appendix:train_para}
The hyperparameters of model training are shown in Table \ref{tab:hyperparameter}.
\begin{table}[h]
\caption{Hyper Parameters of the mdoels.}
\label{tab:hyperparameter}
\vskip 0.15in
\begin{center}
\begin{small}
\begin{sc}
\begin{tabular}{cccc}
        \toprule
        Parameters & Perov\_5 & Carbon\_24 &MP\_20 \\
        \midrule
        Epoch & 3500 &4000 & 1500 \\
        RBF & Bessel & Hybrid & Gaussian \\
        Cutoff & 6 & 6 & 10 \\
        Max Neighbors & 20  & 20  & 50 \\
        Latent Dims & 64& 64  & 64 \\
        lr & 0.0001 & 0.0001 & 0.0001 \\
    \bottomrule
    \end{tabular}
\end{sc}
\end{small}
\end{center}
\vskip -0.1in
\end{table}
\subsection{Loss Function}
\label{appendix:loss_fuc}
The loss function is analogous to that used in CDVAE and can be expressed as follows:
\begin{equation}
\begin{split}
    \mathcal{L}
    & =\mathcal{L}_{Pred}+\mathcal{L}_{Dec}+\mathcal{L}_{KL} \\
    & =\lambda_{A_c}\mathcal{L}_{A_c}+\lambda_{N}\mathcal{L}_{N}+\lambda_{L}\mathcal{L}_{L}+\\
& \space\space\space\space\space    \lambda_{A}\mathcal{L}_{A}+\lambda_{X}\mathcal{L}_{X}+\beta\mathcal{L}_{KL},
\end{split}
\end{equation}
For all three datasets, we use $\lambda_{A_c}=1$, $\lambda_{L}= 10$, $\lambda_{N}=1$, $\lambda_{X}=10$, $\lambda_{A} = 1$. For Perov\_5, MP\_20, we use $\beta=0.01$, and for Carbon\_24, we use $\beta=0.03$.
\subsection{RBF Comparison Results}
\label{appendix:rbf}
\begin{figure}[ht]
\vskip 0.1in
\begin{center}
\centerline{\includegraphics[width=\columnwidth]{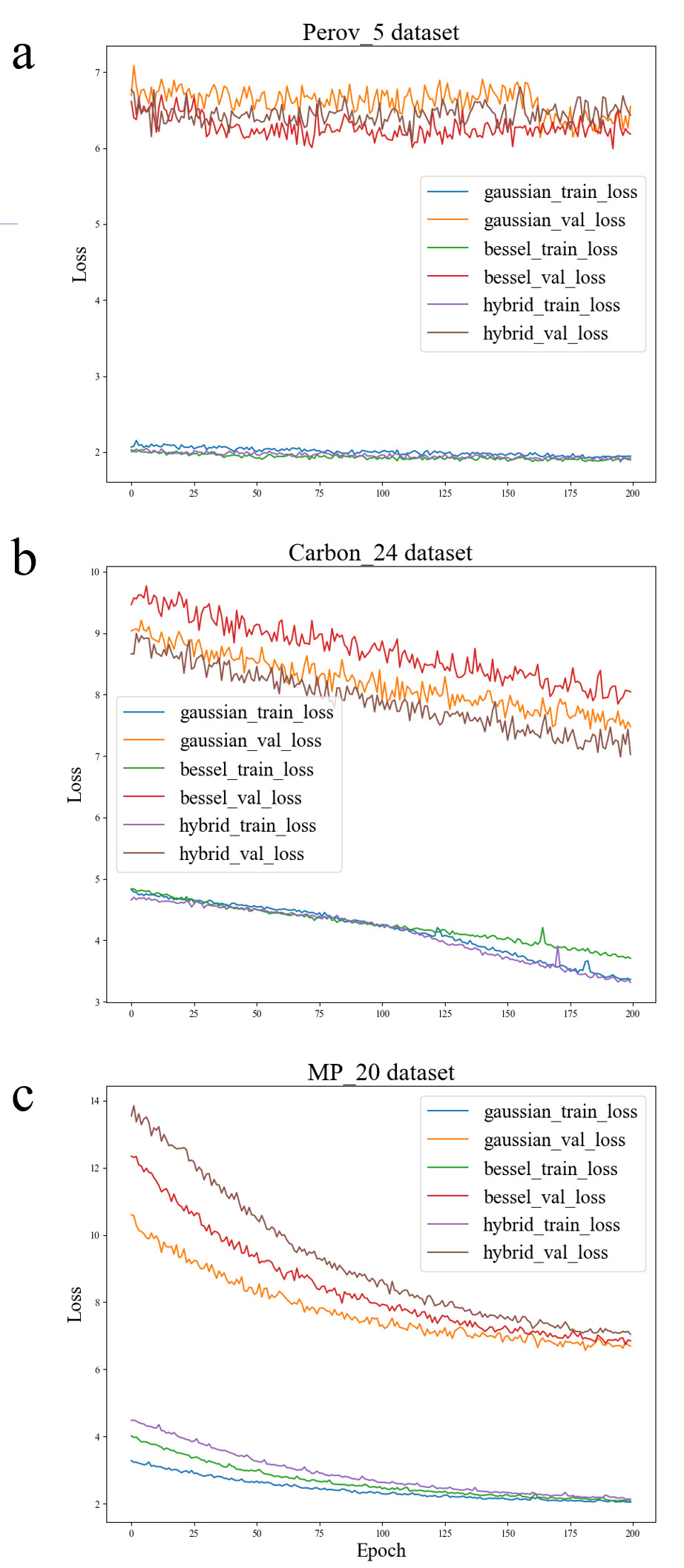}}
\caption{The Comparison of different RBF on Perov\_5, Carbon\_24, MP\_20. Figure a) illustrates the curve generated by training on Perov\_5. It is evident that the model utilizing the Bessel RBF exhibits the most effective training results. Similarly, b) Carbon\_24 - Hybrid RBF, c) MP\_20 - Gaussian RBF.}
\label{rbf_comp}
\end{center}
\vskip -0.1in
\end{figure}

On each of the three datasets—Perov\_5, Carbon\_24, and MP\_20—three models with identical parameters are trained, differing only in the distance expansion function. The control variables include the Gaussian RBF, Bessel RBF, and Hybrid RBF. The convergence curves of the training loss are analyzed, and the results are presented in Fig.\ref{rbf_comp}.

\section{Metrics}
\label{appendix:metric}
\subsection{Validity}
Validity indicators encompass structural validity and component validity. The criterion for structural validity stipulates that as long as the shortest distance between any pair of atoms in the crystal structure exceeds 0.5 \AA, the structure is considered valid. The criterion for component validity requires that the total charge calculated by SMACT\cite{davies2019smact} is neutral, indicating that the combination is valid.
\subsection{Coverage}
Coverage metrics encompass recall and precision. For each true crystal, determine the minimum structural distance $D_{struct}$ and composition distance $D_{comp}$ to all generated crystals. Similarly, for each generated crystal, calculate the minimum structural distance and composition distance to all true crystals. Recall is defined as the proportion of minimum structural and composition distances that fall below specified thresholds - $T_{struct}$ and $T_{comp}$, reflecting the ability of the generated crystals to accurately represent the true crystals in terms of structure and composition. Precision, on the other hand, is the proportion of minimum structural and composition distances that are below the same thresholds, indicating the accuracy of the generated crystals in relation to the true crystals' structure and composition. The detailed settings of thresholds are seen in Table \ref{tab:thresholds}.
\begin{table}[h]
\caption{Coverage Thresholds of various datasets.}
\label{tab:thresholds}
\vskip 0.15in
\begin{center}
\begin{small}
\begin{sc}
\begin{tabular}{ccc}
        \toprule
        Datasets & Structure &  Composition\\
        \midrule
        Perov\_5 & 0.2 & 4.0 \\
        Carbon\_24 & 0.2 & 4.0 \\
        MP\_20 & 0.4 & 10.0 \\
    \bottomrule
    \end{tabular}
\end{sc}
\end{small}
\end{center}
\vskip -0.1in
\end{table}

\subsection{Property statistics}
Property statistics calculate the Earth Mover's Distance (EMD) between the generated material and the ground truth material property distributions, which include density ($\rho$, unit \texttt{$g/cm^{3}$}), energy predicted by an independent Graph Neural Network (GNN) ($E$, unit \texttt{$eV/atom$})\cite{xie2021crystal}, and the number of unique elements.


\end{document}